\documentclass[pre,twocolumn,showpacs,showkeys,preprintnumbers,amsmath,amssymb]{revtex4}

\usepackage{graphicx}
\usepackage{dcolumn}
\usepackage{bm}

\begin{document}


\title{Dynamic phase transition properties and hysteretic behavior of a ferrimagnetic core-shell nanoparticle in the presence of a time dependent magnetic field}

\author{Yusuf Y\"{u}ksel}
\author{Erol Vatansever}
\author{Hamza Polat}\email{hamza.polat@deu.edu.tr}

\affiliation{Department of Physics, Dokuz Eyl\"{u}l University, TR-35160 Izmir, Turkey}
\date{\today}

\begin{abstract}
We have presented dynamic phase transition features and stationary-state behavior of a ferrimagnetic small nanoparticle system with a core-shell structure. By means of detailed Monte Carlo simulations, a complete picture of the phase diagrams and magnetization profiles have been presented and the conditions for the occurrence of a compensation point $T_{comp}$ in the system have been investigated. According to N\'{e}el nomenclature, the magnetization curves of the particle have been found to obey  P-type, N-type and Q-type classification schemes under certain conditions. Much effort has been devoted to investigation of hysteretic response of the particle and we observed the existence of triple hysteresis loop behavior which originates from the existence of a weak ferromagnetic core coupling $J_{c}/J_{sh}$, as well as a strong antiferromagnetic interface exchange interaction $J_{int}/J_{sh}$. Most of the calculations have been performed for a particle in the presence of oscillating fields of very high frequencies and high amplitudes in comparison with exchange interactions which resembles a magnetic system under the influence of ultrafast switching fields. Particular attention has also been paid on the influence of the particle size on the thermal and magnetic properties, as well as magnetic features such as coercivity, remanence and compensation temperature of the particle. We have found that in the presence of ultrafast switching fields,  the particle may exhibit a dynamic phase transition from paramagnetic to a dynamically ordered phase with increasing ferromagnetic shell thickness.
\end{abstract}

\maketitle

\section{Introduction}\label{intro}
In recent years, influences of small-size and surface effects on the magnetic properties of magnetic nanoparticles have
provided a conspicuous and productive field for the interaction between theoretical works \cite{kodama} and technological
\cite{kim}, as well as biomedical applications  \cite{pankhurst, rivas}. As the physical size of a magnetic system
reduces to a characteristic length, surface effects become dominant on the system, hence, some unusual and interesting magnetic
phenomena can be observed, which may differ from those of bulk materials \cite{kodama2}. Recent developments in the experimental
techniques allow the scientists to fabricate such kinds of fine nanoscaled materials \cite{ruhrig, schrefl}, and the magnetization
of certain nanomaterials such as  $\mathrm{\gamma}$-$\mathrm{Fe_{2}O_{3}}$ nanoparticles has been experimentally measured
\cite{martinez}. In particular, magnetic nanowires and nanotubes such as $\mathrm{ZnO}$ \cite{fan}, $\mathrm{FePt}$ and
$\mathrm{Fe_{3}O_{4}}$ \cite{su} can be synthesized by various experimental techniques and they have many applications
in nanotechnology \cite{skomski, schlorb}.

From the theoretical point of view, many studies have been performed regarding the magnetic properties of nanoparticles,
such as nanowire, nanotube, and nanorod systems, and theoretical works can be classified in two basic categories based on
the investigation of equilibrium or nonequilibrium phase transition properties of such nanoscaled magnetic structures.
Namely, in the former group, equilibrium properties of these systems have been investigated by a variety of techniques
such as mean field theory (MFT) \cite{leite, kaneyoshi1, kaneyoshi7, kaneyoshi8},
effective-field theory (EFT) \cite{kaneyoshi3, kaneyoshi7, kaneyoshi8, wang3, keskin, bouhou}, Green functions formalism
\cite{garanin}, variational cumulant expansion (VCE) \cite{wang, wang2} and Monte Carlo (MC) simulations \cite{iglesias, hu1,
iglesias3,  vasilakaki2, zaim, zaim2, jiang, yuksel}. Based on MC simulations, particular attention has been paid on the
exchange bias (EB) effect in magnetic core-shell
nanoparticles where the hysteresis loop exhibits a shift below the N\'{e}el temperature of the antiferromagnetic shell
due to the exchange coupling on the interface region of ferromagnetic core and antiferromagnetic shell. The readers may refer
to \cite{iglesias3} for a rigorous review about the EB phenomena.

It is a well known fact that physical properties of a bulk material are independent from size; however, below a critical size,
nanoparticles often exhibit size-dependent properties, and some unique phenomena have been reported, such as superparamagnetism
\cite{kittel,jacobs}, quantum tunneling of the magnetization \cite{chudnovsky}, and unusual large coercivities \cite{kneller}.
As an example, it has been experimentally shown that $\mathrm{La_{0.67}Ca_{0.33}MnO_{3}}$ (LCMN) nanoparticle exhibits a negative
core-shell coupling, although the bulk LCMN is a ferromagnet \cite{bhowmik1,bhowmik2}. Moreover, as a theoretical example, the total
magnetizations in a nanoscaled transverse Ising thin film with thickness $L$ are investigated by the use of both the EFT with correlations
and MFT, and it has been shown that the magnetization may exhibit two compensation points with the increasing film thickness \cite{kaneyoshi8}.
The phenomena of two compensation points observed in the nanoscaled thin films has also been reported for bulk ferrimagnetic materials
\cite{bobak, cekiz2}. However, the origin of the existence of such a phenomenon  in the nanoscaled magnets is quite different from those observed in the bulk ferrimagnetic materials. Namely, a compensation point originates in the bulk systems due
to the different temperature dependence of the atomic moments of the sublattices \cite{buendia}. However, nanoscaled magnetic particles such
as nanowires or nanotubes exhibit a compensation point, due to the presence of an antiferromagnetic interface coupling between the core and
the shell, even if the lattice sites in the particle core and shell are occupied by identical atomic moments. Hence, theoretical investigation
of ferrimagnetism in nanoparticle systems has opened a new field in the research of the critical phenomena in nanoscaled magnetic particles
\cite{kaneyoshi1}.  According to recent MC studies \cite{zaim, zaim2, jiang, yuksel}, it has been shown that the core-shell morphology can be successfully
applied in equilibrium properties of nanoparticles formed by more than one compound (i.e. ferrimagnetic nanostructures) since the concept is
capable of explaining various characteristic behaviors observed in nanoparticle magnetism. Namely, we learned from these works that
compensation point fairly depends on the particle size. Therefore, nanoscaled magnets such as nanowires, nanotubes, etc. are currently
considered as promising candidates due to their potential utilization as ultra-high density recording media.

On the other hand, a magnetic system exhibits nonequilibrium phase transition properties in the presence of a driving magnetic field.
Namely, when a magnetic material is subject to a periodically varying time dependent magnetic field, the system may not respond to
the external magnetic field instantaneously which causes interesting behaviors due to the competing time scales of the
relaxation behavior of the system and periodic external magnetic field. At high temperatures and for the high amplitudes of the periodic magnetic field,
the system is able to follow the external field with some delay while this is not the case for low temperatures and small magnetic field amplitudes.
This spontaneous symmetry breaking indicates the presence of a dynamic phase transition (DPT) \cite{tome, lo, chakrabarti} which shows itself in the
dynamic order parameter (DOP) which is defined as the time average of the magnetization over a full period of the oscillating field. Related to this nonequilibrium phenomena, in recent years, based on Glauber type of stochastic dynamics \cite{glauber}, a few theoretical studies have been devoted to the investigation of dynamical aspects of phase transition properties of cylindrical Ising nanowire and nanotube systems in the presence of a
time-dependent magnetic field within the EFT with correlations \cite{deviren1, deviren2}. In those studies, the authors analyzed the  temperature
dependencies of the dynamic magnetization, hysteresis loop area and dynamic correlation between time dependent magnetization and magnetic field, and it has been reported that dynamic magnetization curves can be classified into well known categories, according to N\'{e}el theory of ferrimagnetism \cite{neel,strecka}. Furthermore, based on MC simulations and by using unixally anisotropic Heisenberg model, frequency dispersion of dynamic hysteresis in a
core-shell magnetic nanoparticle system has been studied by Wu et al. \cite{wu}, in order to determine whether the dynamic hysteresis loops
obey the power-law scaling or not, and they concluded that the frequency dispersion of the dynamic hysteresis shows both the spin-reversal and spin-tilting resonances, and also they found that the exchange coupling on the core-shell interface has no effect on the power-law scaling of the dynamic hysteresis dispersion.

As can be seen in the previously published works mentioned above, equilibrium phase transition properties of nanoparticle systems have been almost
completely understood, whereas nonequilibrium counterparts needs particular attention and the following questions need to be answered: (i) What is
the effect of the amplitude and frequency of the oscillating magnetic field on the dynamic phase transition properties (i.e. critical and compensation temperatures) of the nanoparticle systems? (ii) What kind of physical relationships exist between the magnetic properties (compensation point and coercivity) of the particle and the system size? Main motivation of the present paper is to attempt to clarify the physical facts underlying these questions. Particular emphasis has also been devoted for the treatment of highly nonequilibrium situation where the applied field amplitudes are of the same order as (or greater than) the exchange interactions, and oscillation periods with tens of time steps (in terms of MC steps per site). The physical realization of this situation can be probably achieved by applying ultrafast laser fields \cite{daniel, sabareesan, qi, sukhov, stanciu, choi} or by using some certain novel materials with low exchange interaction, in comparison with external field strength.

Outline of the paper is as follows: In Section \ref{formulation} we briefly present our model. The results and discussions are presented in Section \ref{results}, and finally Section \ref{conclude} contains our conclusions.

\section{Formulation}\label{formulation}
We consider a cubic ferrimagnetic nanoparticle composed of a spin-3/2 ferromagnetic core which is surrounded by a spin-1 ferromagnetic shell layer.
At the interface, we define an antiferromagnetic interaction between core and shell spins (see figure 1 in Ref. \cite{yuksel}). Construction of such kind of model allows us to simulate a ferrimagnetic small particle formed by more than one compound. The particle is subjected to a periodically
oscillating magnetic field. The time dependent Hamiltonian describing our model of magnetic system can be written as
\begin{eqnarray}\label{eq1}
\nonumber
\mathcal{H}&=&-J_{int}\sum_{<ik>}\sigma_{i}S_{k}-J_{c}\sum_{<ij>}\sigma_{i}\sigma_{j}-J_{sh}\sum_{<kl>}S_{k}S_{l}\\
&&-h(t)\left(\sum_{i}\sigma_{i}+\sum_{k}S_{k}\right),
\end{eqnarray}
where $\sigma=\pm3/2,\pm1/2$ and $S=\pm1,0$ are spin variables in the core and shell sublattices. $J_{int}$, $J_{c}$ and $J_{sh}$ define antiferromagnetic
interface and ferromagnetic core and shell exchange interactions, respectively. $h(t)=h\sin (\omega t)$ represents the oscillating magnetic field,
where $h$ and $\omega$ are the amplitude and the angular frequency of the applied field, respectively. Period of the oscillating magnetic
field is given by $\tau=2\pi /\omega$. $\langle ...\rangle$ denotes the nearest neighbor interactions on the lattice. We fixed the value of $J_{sh}$ to unity throughout
the simulations, and we also use normalized the exchange interactions with $J_{sh}$. Accordingly, amplitude of the oscillating magnetic field has been normalized as $h_{0}=h/J_{sh}$ in the calculations.

In order to simulate the system, we employ Metropolis MC simulation algorithm \cite{binder} to equation (\ref{eq1}) on an $L\times L\times L$ simple-cubic lattice with free boundary conditions (FBC) which is an appropriate choice for such a finite small system. Configurations were generated by selecting the sites in sequence through the lattice and making single-spin-flip attempts, which were accepted or rejected according to the Metropolis algorithm, and $L^{3}$ sites are visited at each time step (a time step is defined as a MC step per site or simply MCS). The frequency $f$ of the oscillating magnetic field is defined in terms of MCS in such a way that
\begin{equation}\label{eq2}
f=\frac{1}{\kappa\Theta_{s}},
\end{equation}
where $\kappa$ is the number of MCSs necessary for one complete cycle of the oscillating field and $\Theta_{s}$ is the time interval. In our simulations,
we choose $\Theta_{s}=1$, hence we get $\tau=\kappa$. Data were generated over $50-100$ independent sample realizations by running most of the simulations for $25 000$
Monte Carlo steps per site after discarding the first $5000$ steps. This amount of transient steps is found to be sufficient for thermalization for almost the whole range of the parameter sets. However, for evaluating the hysteresis loops, in order to guarantee to obtain stable loops, the first few cycles of the external field are considered as transient regime, and after this transient regime, statistical averaging has been performed (see Section \ref{results2}). Error bars were calculated by using the jackknife method \cite{newman}.

Our program calculates the instantaneous values of the core and shell layer magnetizations $M_{c}$ and $M_{sh}$, and the total magnetization $M_{T}$ at the time $t$.
These quantities are defined as
\begin{eqnarray}\label{eq3}
\nonumber
M_{c}(t)=\frac{1}{N_{c}}\sum_{i=1}^{N_{c}}\sigma_{i},\quad M_{sh}(t)=\frac{1}{N_{sh}}\sum_{i=1}^{N_{sh}}S_{i},\\
M_{T}(t)=\frac{N_{c}M_{c}(t)+N_{sh}M_{sh}(t)}{N_{c}+N_{sh}},\ \ \ \ \ \ \ \ \ \ \ \ \ \
\end{eqnarray}
where $N_{c}$ and $N_{sh}$ denote the number of spins in core and shell layers, respectively. From the instantaneous magnetizations, we obtain the dynamic order
parameters as follows
\begin{eqnarray}\label{eq4}
\nonumber
Q_{c}=\frac{\omega}{2\pi}\oint M_{c}(t)dt,\quad Q_{sh}=\frac{\omega}{2\pi}\oint M_{sh}(t)dt,&&\\
Q=\frac{\omega}{2\pi}\oint M_{T}(t)dt,\ \ \ \ \ \ \ \ \ \ \ \ \ \ \ \ \ \ \ \ \ \ \ \ \ \ \ \ \ \ \ \ \
\end{eqnarray}
where $Q_{c}$, $Q_{sh}$ and $Q$ denote the dynamic order parameters corresponding to the core and shell layers, and to the overall lattice ($Q$ is assumed to represent the time averaged total magnetization over a full cycle of the oscillating field), respectively. We also calculate the time average of the total energy of the particle including both cooperative and field parts over a full cycle of the magnetic field
as follows \cite{acharyya}
\begin{eqnarray}\label{eq5}
\nonumber
E_{tot}&=&-\frac{\omega}{2\pi L^{3}}\oint \left[J_{int}\sum_{<ik>}\sigma_{i}S_{k}+J_{c}\sum_{<ij>}\sigma_{i}\sigma_{j}\right.\\
\nonumber
&&+J_{sh}\sum_{<kl>}S_{k}S_{l}+\left.h(t)\left(\sum_{i}\sigma_{i}+\sum_{k}S_{k}\right)\right]dt.\\
\end{eqnarray}
Thus, the specific heat of the system is defined as
\begin{equation}\label{eq6}
C=\frac{dE_{tot}}{dT},
\end{equation}
where $T$ represents the temperature. To determine the dynamic compensation temperature $T_{comp}$ from the computed magnetization data,
the intersection point of the absolute values of the dynamic core and shell magnetizations were found using
\begin{equation}\label{eq7}
|N_{c}Q_{c}(T_{comp})|=|N_{sh}Q_{sh}(T_{comp})|,
\end{equation}
\begin{equation}\label{eq8}
\mathrm{sign}(N_{c}Q_{c}(T_{comp}))=-\mathrm{sign}(N_{sh}Q_{sh}(T_{comp}))
\end{equation}
with $T_{comp}<T_{c}$, where $T_{c}$ is the critical temperature i.e. N\'{e}el temperature. Equations (\ref{eq7}) and (\ref{eq8}) indicate that the signs of the dynamic core and shell magnetizations are different, however, absolute values of them are equal to each other at the compensation point. Hence, in order to characterize the compensation points, we also define two additional order parameters belonging to core and shell layers of the particle as follows:
\begin{equation}\label{eq9}
O_{c}=\frac{N_{c}}{N}Q_{c},\quad O_{sh}=\frac{N_{sh}}{N}Q_{sh}.
\end{equation}

\section{Results and Discussion}\label{results}
In this section, we will focus our attention on the dynamic phase transition properties of the ferrimagnetic nanoparticle system. This section is divided into three parts as follows: In Section \ref{results1}, we have examined the dependence of the critical temperature $T_{c}$ of the particle on the amplitude and frequency of the oscillating magnetic field, as well as the exchange couplings defined in equation (\ref{eq1}). In this section, we have also investigated the conditions for the occurrence of a compensation point $T_{comp}$ in the system. Hysteretic response of the particle to the periodically oscillating magnetic fields have been investigated in Section \ref{results2}, and size dependence of the magnetic properties have been analyzed in Section \ref{results3}. In order to make a comparison with the previously published works where the equilibrium properties of the present system were discussed, we select the number of core and shell spins as $N_{c}=11^3$ and $N_{sh}=L^3-11^{3}$ (the same values as in \cite{yuksel} where $L=15$ is the linear dimension of the lattice) in Sections \ref{results1} and \ref{results2}.
\subsection{Dynamic phase transition features of the particle}\label{results1}
\begin{figure}\begin{center}
\includegraphics[width=8cm]{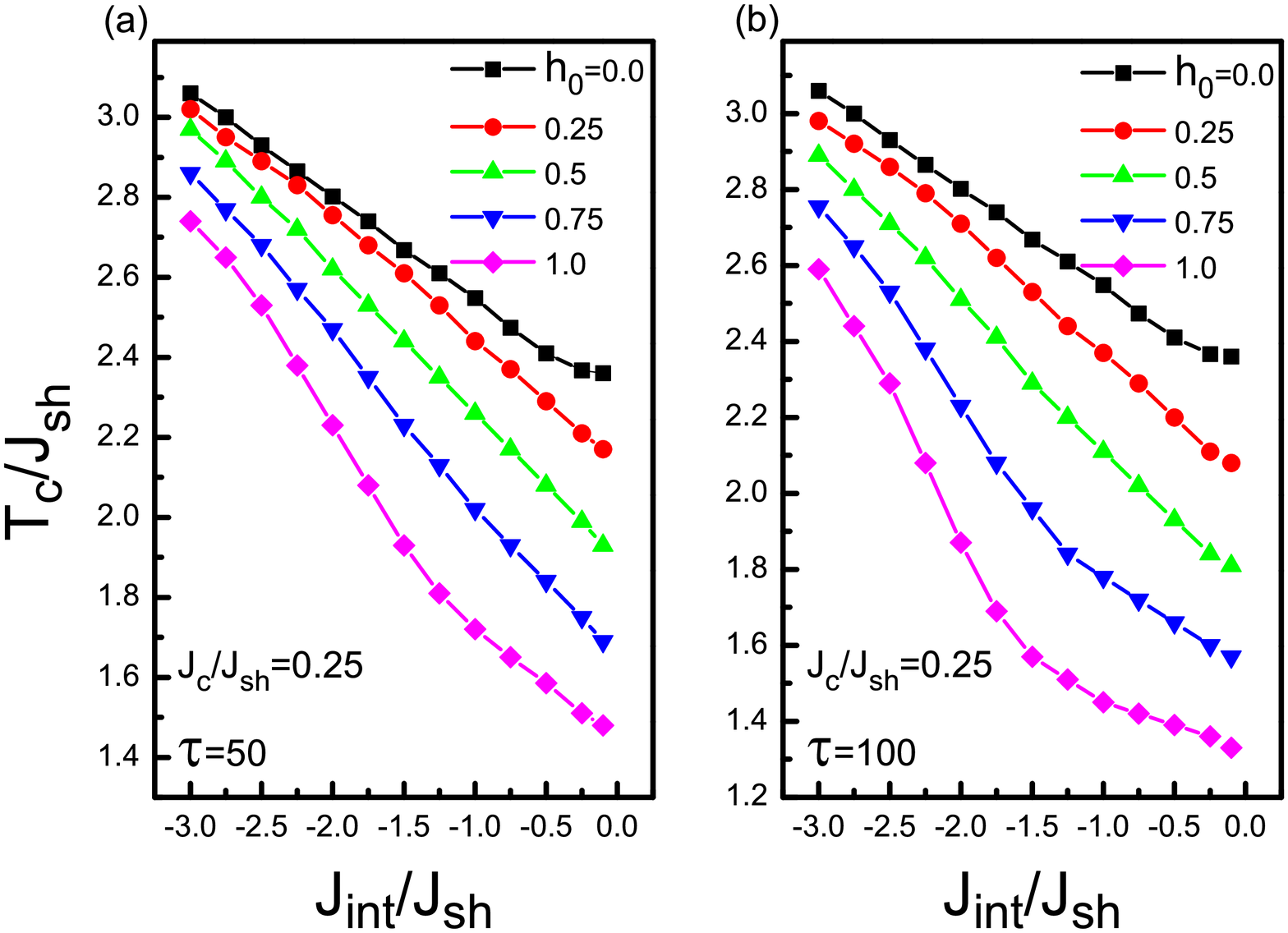}
\includegraphics[width=3.5cm]{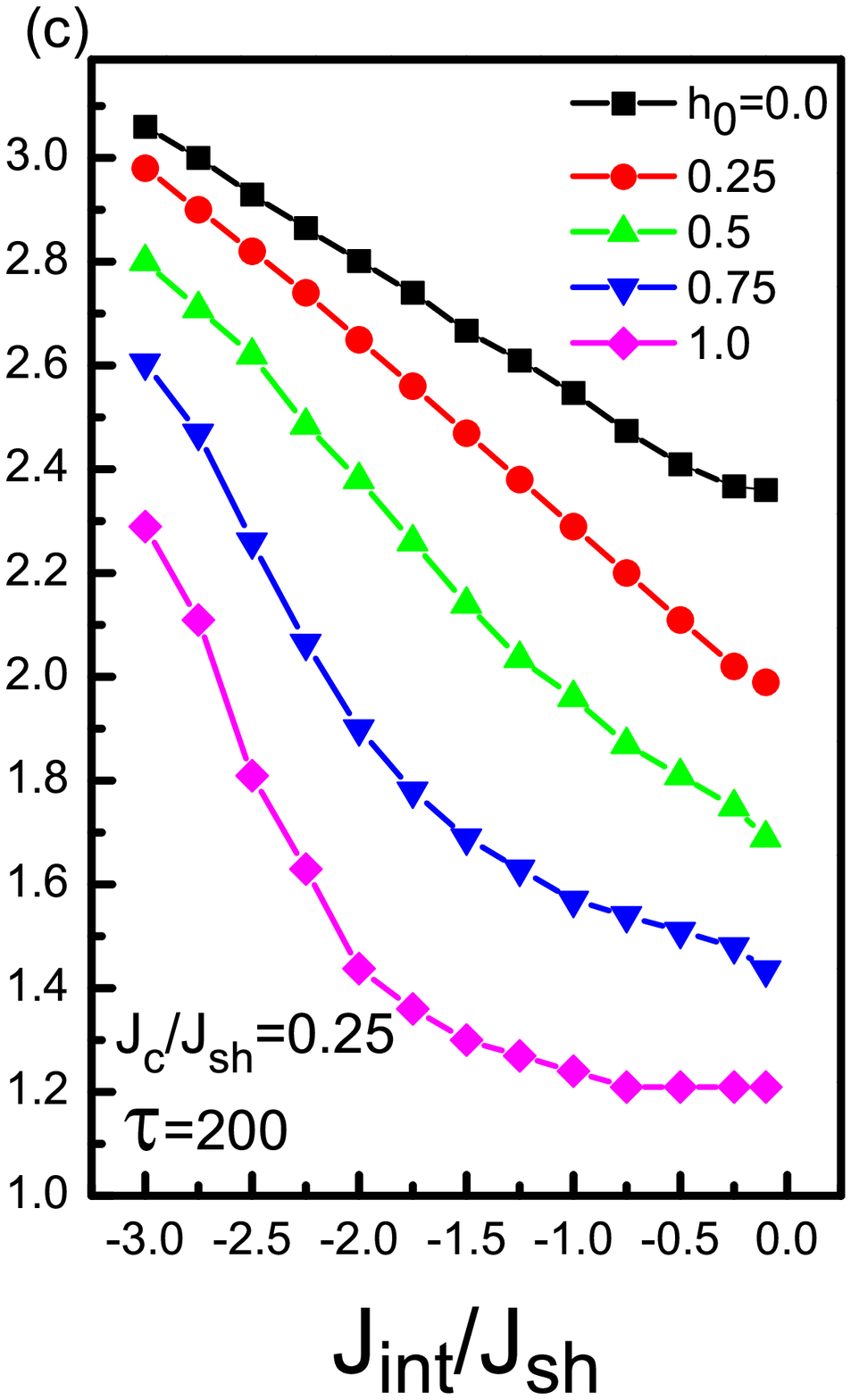}\\
\caption{Phase diagrams of the system in $(T_{c}/J_{sh}-J_{int}/J_{sh})$ plane for $J_{c}/J_{sh}=0.25$ with some selected values of the external field amplitude $h_{0}=0.0,0.25,0.50,0.75,1.0$. The curves are plotted for three values of oscillation period: (a) $\tau=50$, (b) $\tau=100$, and (c) $\tau=200$.}\label{fig1}
\end{center}
\end{figure}
\begin{figure}\begin{center}
\includegraphics[width=8cm]{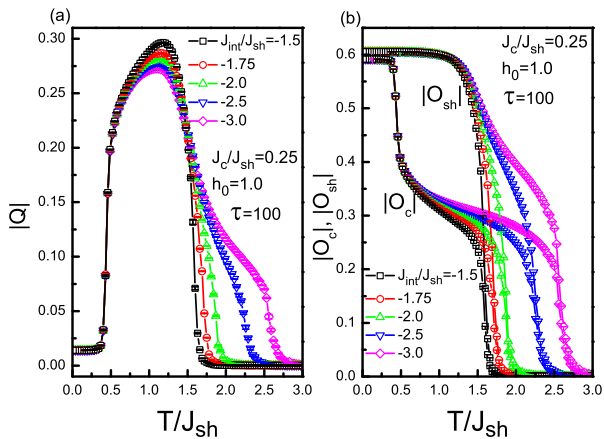}\\
\caption{Effect of the antiferromagnetic interface coupling between core and shell spins on the temperature dependencies of order parameters $Q$, $O_{c}$, and $O_{sh}$ for a combination of Hamiltonian parameters corresponding to phase diagrams depicted in figure \ref{fig1}.}\label{fig2}
\end{center}
\end{figure}

In order to clarify the influence of antiferromagnetic interface coupling $J_{int}/J_{sh}$ between core and shell layers on the dynamic phase transition properties of the system, we represent the phase diagrams in a $(T_{c}/J_{sh}-J_{int}/J_{sh})$ plane with three oscillation period values $\tau=50,100,200$ which realizes ultra-fast switching external fields, and for some selected values of the field amplitude $h_{0}$ in figures \ref{fig1}a-\ref{fig1}c. Here, we consider a weak ferromagnetic interaction, such as $J_{c}/J_{sh}=0.25$ for the core spins which simulates a surface exchange enhancement in the system. One of the common findings in these figures is that transition temperature $T_{c}$ values gradually increase as the strength of the antiferromagnetic interface coupling $J_{int}/J_{sh}$ increases. This phenomenon is independent from amplitude $h_{0}$ and period $\tau$ of the oscillating magnetic field. At high oscillation period values (i.e. at relatively low frequencies), dynamic magnetization $M_{T}$ corresponding to the instantaneous ferrimagnetic order parameter of the particle can respond to the oscillating magnetic field with some delay whereas as the period of the external magnetic field gets lower, a competition occurs between the period $\tau$ of the field and the relaxation time of the system, hence the dynamic magnetization cannot respond to the external field due to the increasing phase lag between the field and the magnetization $M_{T}$. As a result, this makes the occurrence of the dynamic phase transition difficult. In addition, for weak $|J_{int}|$ values, core and shell layers of the particle become independent of each other. As the strength of the antiferromagnetic interface interaction gets increased then it becomes dominant against the periodic local fields, and the particle exhibits a strong ferrimagnetic order. Hence, a relatively large amount of thermal energy is needed to observe a dynamic phase transition in the system, due to the response of the spins to the external magnetic field. As the value of the field amplitude increases then the antiferromagnetic exchange interaction $J_{int}$ loses its dominance against the external field amplitude and it becomes possible to observe a dynamic phase transition at lower temperatures. Consequently, ferrimagnetically ordered phase region in the phase diagrams shown in figures \ref{fig1}a-\ref{fig1}c gets narrower with increasing $h_{0}$ and $\tau$ values. In figures \ref{fig2}a and \ref{fig2}b, we depict the effect of the antiferromagnetic interface coupling $J_{int}/J_{sh}$ on the temperature dependencies of dynamic order parameters, corresponding to the phase diagrams shown in figure \ref{fig1}b. As an interesting observation, we can see from figure \ref{fig2} that although the ferromagnetic exchange coupling of the particle core is relatively weaker than that of the shell layer (i.e. $J_{c}/J_{sh}=0.25$), both the core and shell layers undergo a dynamic phase transition at the same critical temperature which is a result of the relatively strong interface coupling $J_{int}$. As seen in the magnetization curves shown in figure \ref{fig2}a, magnetization of the present nanoparticle system can exhibit similar features as observed in the bulk ferrimagnetic systems. In the bulk ferrimagnetism of N\'{e}el \cite{neel,strecka}, it is possible to classify the thermal variation of the total magnetization curves in some certain categories. According to this nomenclature, the system exhibits P-type behavior at which the magnetization shows a temperature-induced maximum with increasing temperature. At this point, we should note that this result conflicts with some recent works \cite{zaim, zaim2, yuksel}. In particular,  in these studies the existence of at least one compensation point has been predicted for equilibrium properties of the system, i.e. for $h_{0}=0$. However, the total magnetization of the system has been defined in such a way that the inequality of the number of spins in the core and shell layers has been ignored in the aforementioned works which completely effects the results in qualitative manner. The situation can be clarified by analyzing core and shell magnetizations of the particle. These results are given in figure \ref{fig2}b. As seen in this figure, core (with 1331 spins) and shell (with 2044 spins) magnetizations do not cancel each other, hence  we cannot observe any compensation point in the system for given set of system parameters. Accordingly, we can conclude that in the presence of surface exchange enhancement (such as $J_{c}/J_{sh}=0.25$) and at high oscillation frequencies the system does not exhibit compensation phenomena.

\begin{figure}\begin{center}
\includegraphics[width=8cm]{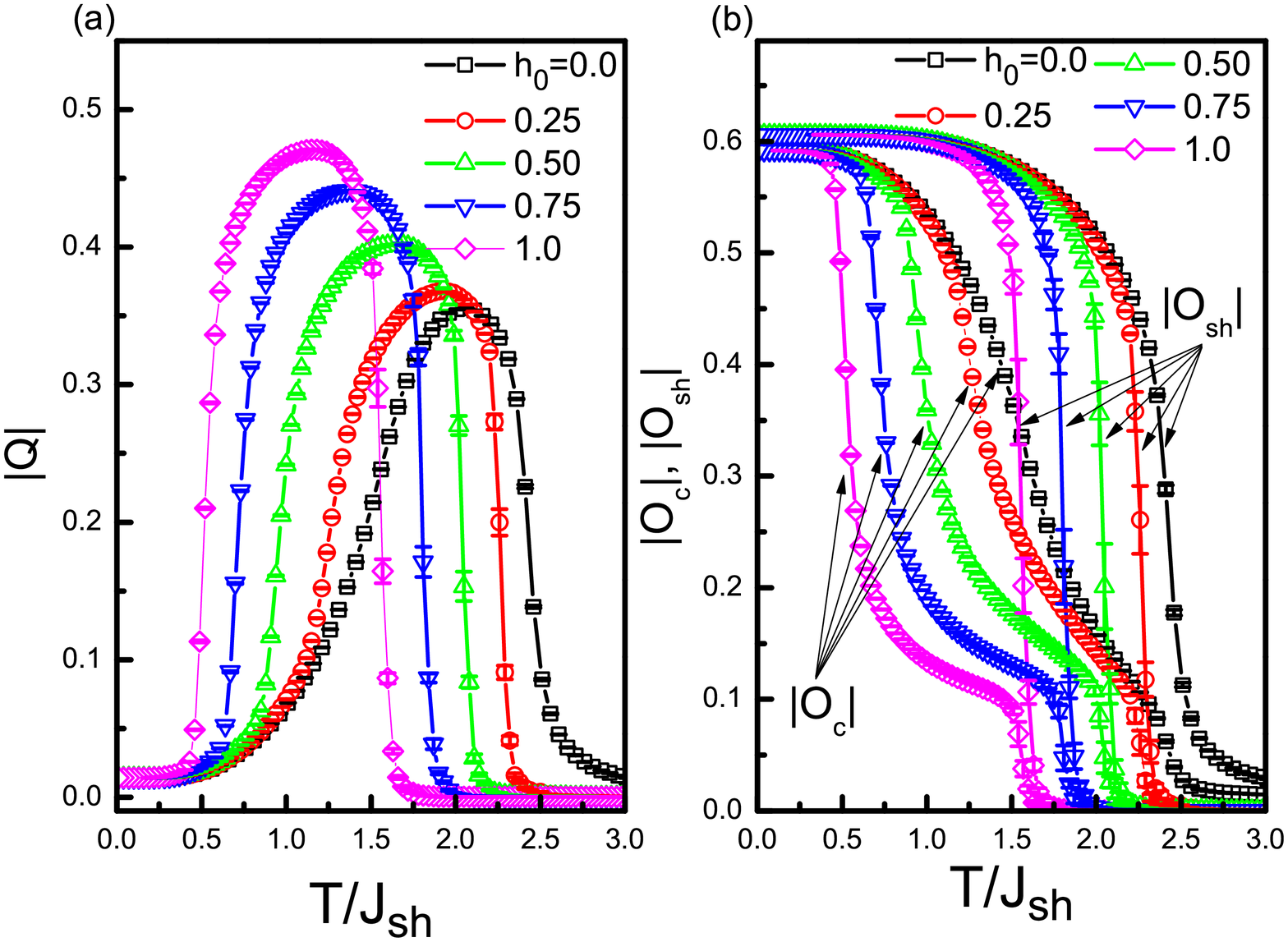}
\includegraphics[width=4.0cm]{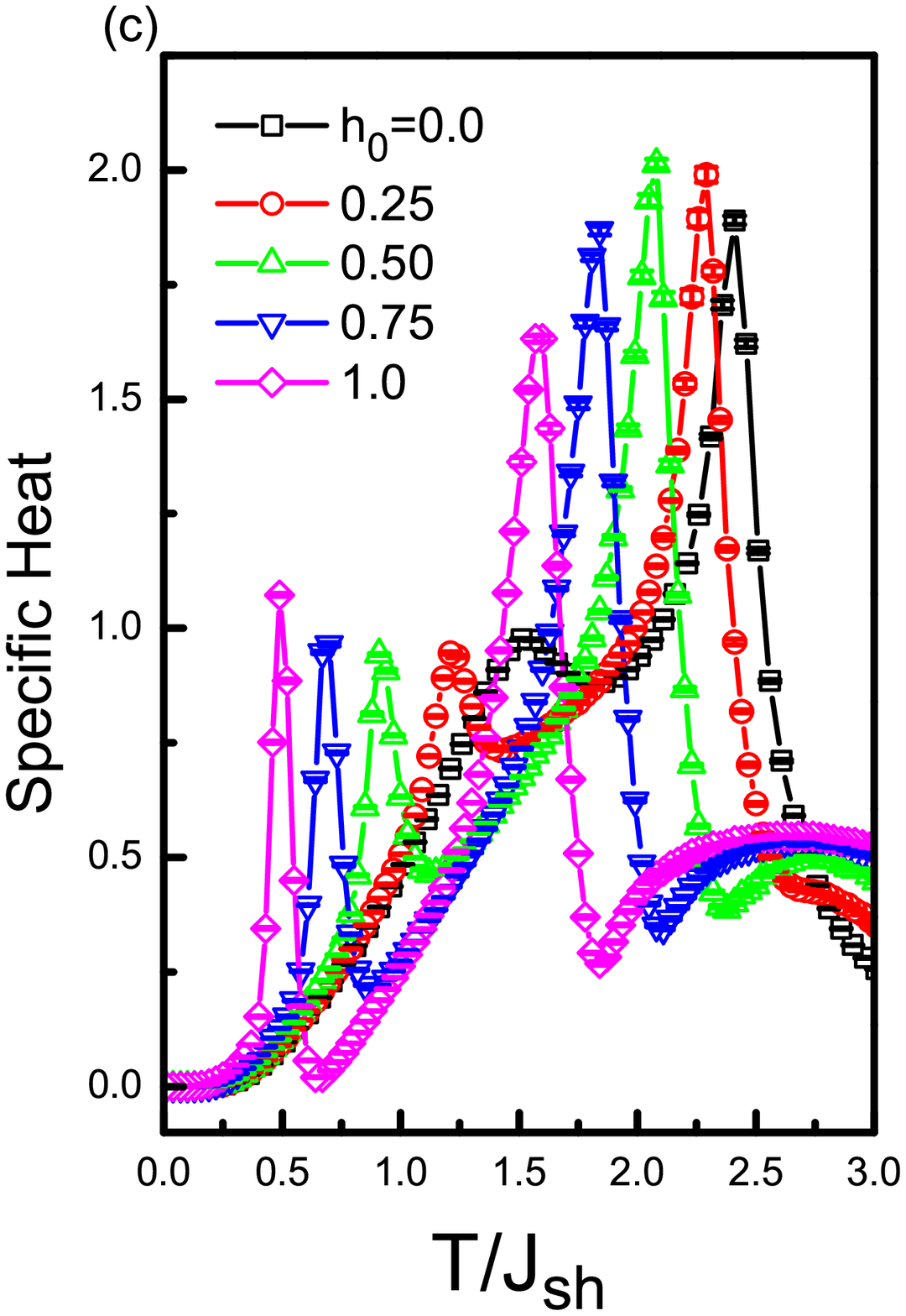}\\
\caption{Effect of the external field amplitude $h_{0}$ on the temperature dependencies of (a) total magnetization $Q$, (b) dynamic order parameters $O_{c}$, $O_{sh}$, and (c) dynamic heat capacity $C$ for $J_{int}/J_{sh}=-0.5$, $J_{c}/J_{sh}=0.25$, and $\tau=50$ with $h_{0}=0.0,0.25,0.50,0.75,1.0$.}\label{fig3}
\end{center}
\end{figure}
\begin{figure}\begin{center}
\includegraphics[width=8cm]{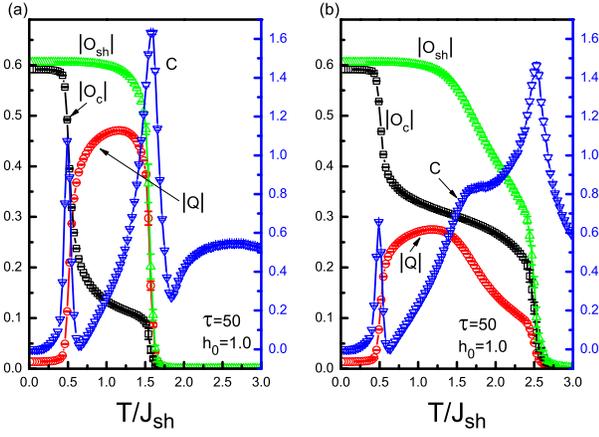}\\
\caption{Temperature dependencies of total magnetization $Q$, dynamic order parameters $O_{c}$, $O_{sh}$, and specific heat $C$ of the particle for $J_{c}/J_{sh}=0.25$, $h_{0}=1.0$, and $\tau=50$ with (a) $J_{int}/J_{sh}=-0.5$ and (b) $J_{int}/J_{sh}=-2.5$. }\label{fig4}
\end{center}
\end{figure}

Next, in figure \ref{fig3}, we present the influence of the amplitude $h_{0}$ of the external field on the temperature dependencies of total magnetization $Q$, dynamic order parameters $O_{c}$ and $O_{sh}$, as well as the dynamic heat capacity curves of the particle, corresponding to the phase diagrams depicted in figure \ref{fig1}a. In figure \ref{fig3}a, total magnetization curves of the overall system are plotted. As seen in figure \ref{fig3}a, magnetization curves exhibit P-type behavior, and $T_{c}$ values decrease with increasing $h_{0}$ values. On the other hand, dynamic heat capacity curves which are depicted in figure \ref{fig3}c exhibit a hump at lower temperatures and a sharp peak at the transition temperature. Schottky-like rounded humps observed in the heat capacity curves get sharper with increasing $h_{0}$ which is a result of a sudden change in the core magnetization. Behavior of the dynamic specific heat curves corresponding to highly non-monotonous magnetization profiles can be better observed in figure \ref{fig4}. Namely, when the magnetization curves exhibit conventional P-type behavior, dynamic specific heat curves exhibit a hump, and a sharp peak whereas if the magnetization shows P-type behavior with two separate maxima at $T<T_{c}$ then the specific heat curves exhibit two distinct Schottky-like rounded humps, and a sharp transition peak. When the amplitude $h_{0}$ of the external field is sufficiently large, then the first hump observed at low temperatures gets sharper whereas the shape of the other one which is observed at higher temperatures does not change, since it originates from the thermal variation of the magnetization of the shell layer which does not exhibit a sudden variation as the temperature is varied.

\begin{figure}\begin{center}
\includegraphics[width=8cm]{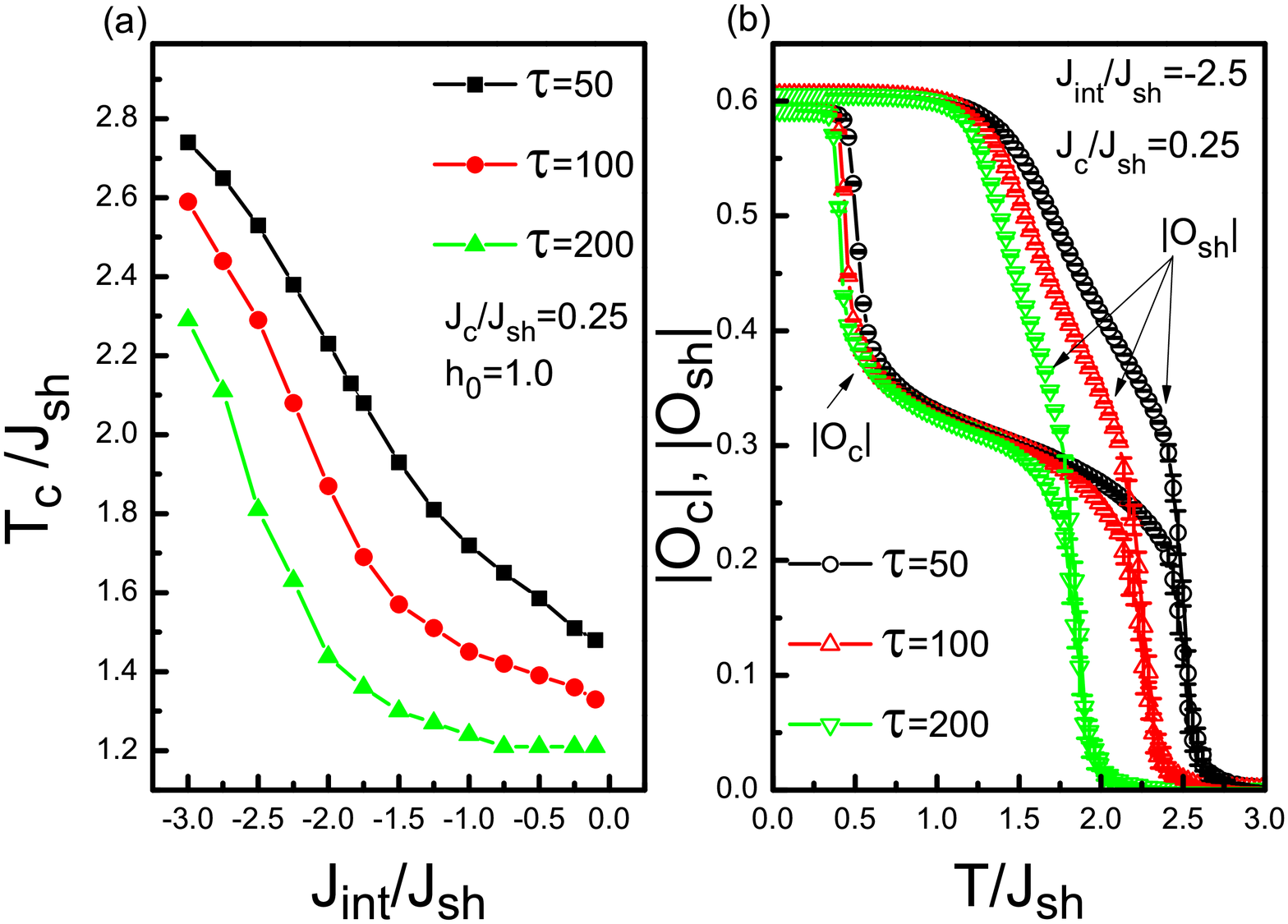}\\
\caption{(a) Phase diagrams of the particle in $(T_{c}/J_{sh}-J_{int}/J_{sh})$ plane for $J_{c}/J_{sh}=0.25$, $h_{0}=1.0$, and $\tau=50,100,200$. (b) The effect of the oscillation period $\tau$ on the temperature dependencies of dynamic order parameters $O_{c}$ and $O_{sh}$ of core and shell layers of the particle for $J_{int}/J_{sh}=-2.5$, $J_{c}/_{sh}=0.25$, $h_{0}=1.0$, and $\tau=50,100,200$.  }\label{fig5}
\end{center}
\end{figure}

In figure \ref{fig5}, we examine the effect of the external field period $\tau$ on the dynamical phase transition characteristics of the particle. Phase diagrams in this figure are plotted for a value of the field amplitude $h_{0}/J_{sh}=1.0$ with relatively high frequencies (in comparison with number of mcss) in which nonequilibrium properties of the system have been simulated under ultrahigh fields and ultra fast speeds. From figure \ref{fig5}a,  one can clearly observe that $T_{c}$ values are depressed with increasing $\tau$. The physical facts underlying the behaviors observed in figure \ref{fig5}a are identical to those emphasized in figure \ref{fig2}. Therefore we will not discuss these interpretations here. However, as a complementary investigation, let us represent certain magnetization profiles corresponding to the phase diagrams given in figure \ref{fig5}a. For instance, as seen in figure \ref{fig5}b, when the antiferromagnetic exchange interaction strength is selected as $J_{int}/J_{sh}=-2.5$, dynamic order parameters $O_{c}$ and $O_{sh}$ of the core and shell layers never intersect each other, which causes the observation of P-type characteristics in total magnetization curves. We have also performed simulations for the case of $J_{int}/J_{sh}=-0.5$, but we have not observed any significant difference in characteristic behavior of magnetization curves. Hence, based on the results given in figures \ref{fig4} and \ref{fig5}, we see that the existence of strong antiferromagnetic exchange interaction (such as $J_{int}/J_{sh}=-2.5$) is not sufficient for the occurrence of compensation phenomena in the system.

\begin{figure}\begin{center}
\includegraphics[width=8cm]{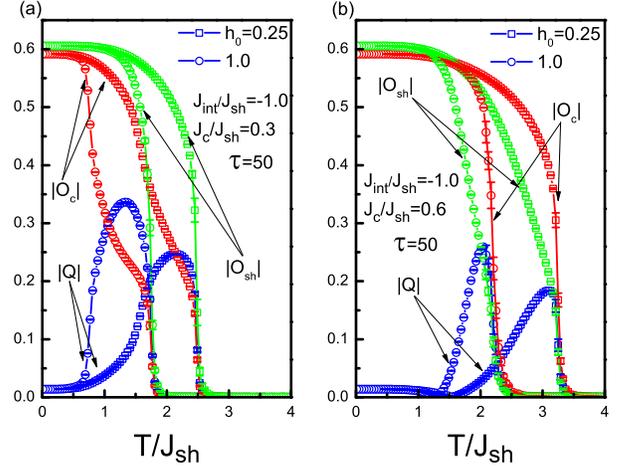}\\
\caption{Effect of $J_{c}/J_{sh}$ on the temperature dependencies of $Q$, $O_{c}$, and $O_{sh}$ curves of the system for $J_{int}/J_{sh}=-1.0$, $\tau=50$, and $h_{0}=0.25,1.0$ with (a) $J_{c}/J_{sh}=0.3$, (b) $J_{c}/J_{sh}=0.6$. }\label{fig6}
\end{center}
\end{figure}

As a final investigation of this subsection, we will discuss the influence of the ferromagnetic exchange interaction of the core layer $J_{c}/J_{sh}$ on the magnetization profiles of the particle in figure \ref{fig6}. At first sight, according to figure \ref{fig6}, we can clearly claim that as the value of $J_{c}/J_{sh}$ increases then the transition temperature of the system also increases. We can also mention that the phase transition temperature of the particle shell is directly related to the value of $J_{c}/J_{sh}$, since the antiferromagnetic interface interaction is relatively large as $J_{int}/J_{sh}=-1.0$. Furthermore, for $J_{c}/J_{sh}=0.3$, the magnetization of the particle exhibits P-type behavior, whereas for $J_{c}/J_{sh}=0.6$, a compensation temperature appears which decreases with increasing $h_{0}$, and we find an N-type dependence at which the magnetization is characterized by a compensation point at which total magnetization $Q$ reduces to zero due to the complete cancelation of the core and shell layer magnetizations. These observations are represented in figures \ref{fig6}a and \ref{fig6}b, respectively.

\subsection{Hysteretic response of the particle to the periodically oscillating magnetic fields}\label{results2}
Hysteresis behavior in magnetic systems originates in response to varying magnetic fields, and it is one of the most important features of real magnetic materials. In dynamic systems, the phenomenon occurs as a result of a dynamic phase lag between instantaneous magnetization and periodic external magnetic field. In contrast to the behavior observed in static models where the strength of the external field does not change with time explicitly, dynamic hysteresis in nonequilibrium phase transitions is characterized by a dynamic symmetry loss at high oscillation frequencies of the external field. The shape of a hysteresis loop is determined by the coercivity and remanent magnetization of the magnetic material. In particular, coercivity -which is defined as the required amount of the external magnetic field to reduce the magnetization of a material to zero- is an essential physical property of magnetic materials which has a significant importance in technological applications. Moreover, it is worth to note that hysteresis loops of equilibrium systems exhibit coercivity in ferromagnetic phase \cite{sariyer} whereas coercive fields in nonequilibrium systems driven by an oscillating field are always observed in the dynamic paramagnetic phase.
\begin{figure}\begin{center}
\includegraphics[width=8cm]{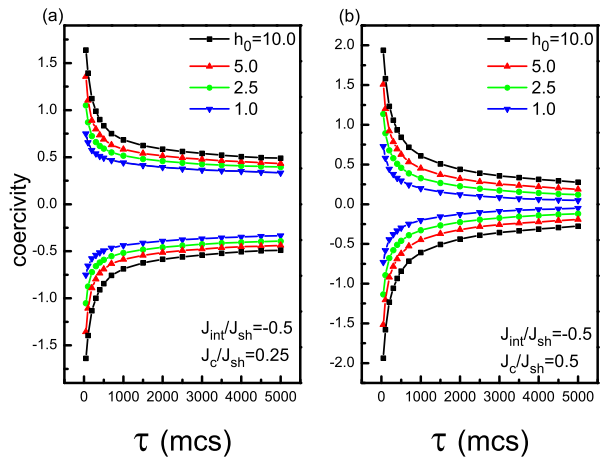}\\
\caption{Variation of the coercive field as a function of the oscillation period $\tau$ for several values of $h_{0}$ with $J_{int}/J_{sh}=-0.5$, and (a) $J_{c}/J_{sh}=0.25$, (b) $J_{c}/J_{sh}=0.5$, respectively.}\label{fig7}
\end{center}
\end{figure}

\begin{figure}\begin{center}
\includegraphics[width=8cm]{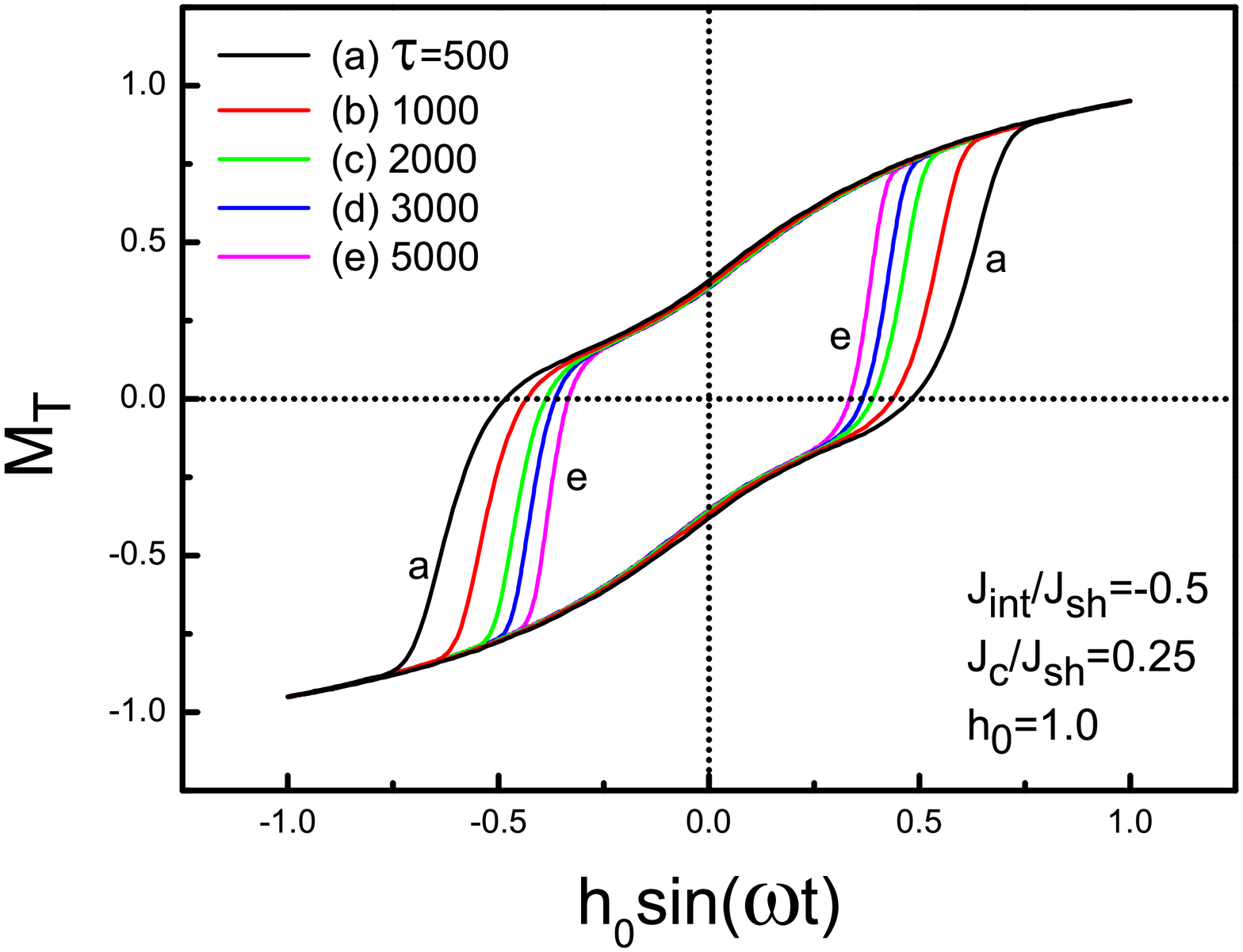}\\
\caption{Hysteresis loops of the particle with a relatively weak $J_{int}/J_{sh}$ value for several values of oscillation period $\tau$ with $J_{int}/J_{sh}=-0.5$, $J_{c}/J_{sh}=0.25$, and $h_{0}=1.0$. The letters accompanying each curve denote the value of $\tau$. }\label{fig8}
\end{center}
\end{figure}

\begin{figure}\begin{center}
\includegraphics[width=8cm]{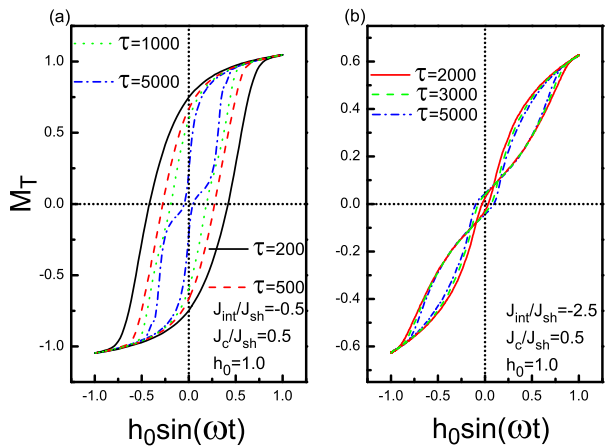}\\
\caption{Hysteresis loops of the particle with some selected values of $\tau$ where $J_{c}/J_{sh}=0.5$, and $h_{0}=1.0$. Effect of antiferromagnetic interface coupling is depicted in (a) for $J_{int}/J_{sh}=-0.5$, and in (b) for $J_{int}/J_{sh}=-2.5$. }\label{fig9}
\end{center}
\end{figure}

Oscillation period and external field amplitude dependence of the coercivity of the particle and dynamic hysteresis loops are shown in figures \ref{fig7}-\ref{fig9} which have been calculated at a temperature $T=0.8T_{c}^{0}$ where $T_{c}^{0}$ is the transition temperature of the system in the absence of the external field. This choice of the temperature allows the system to undergo a purely mechanical phase transition (i.e. magnetic field induced transition). In order to acquire a stationary behavior, the first 100 cycles of the external field has been discarded and the data were collected for 400 cycles.

From figure \ref{fig7}a, for weak ferromagnetic core coupling values such as $J_{c}/J_{sh}=0.25$, coercivity curves exhibit sudden variation with increasing $\tau$ whereas for sufficiently high $\tau$ they exhibit a stable profile. Moreover, it is clear from figure \ref{fig7}a that at low oscillation periods, higher amplitude values mean large coercive fields, however at high $\tau$ values, coercivity becomes independent from $h_{0}$. However, according to figure \ref{fig7}b, as the strength of ferromagnetic exchange interaction $J_{c}/J_{sh}$ between the core spins increases, coercivity may reduce to zero at high $\tau$ values for relatively small $h_{0}$ values. Some typical examples of the dynamic hysteresis loops obtained from the time evolution of instantaneous magnetization $M_{T}$ corresponding to figure \ref{fig7} (where there exists a relatively weak interface interaction such as $J_{int}/J_{sh}=-0.5$ between core and shell layers) are depicted in figure \ref{fig8} for some selected values of oscillation period $\tau$ with $h_{0}=1.0$. From figure \ref{fig8}, we see that remanent magnetization values of the system do not change as $\tau$ varies. On the other hand, width of the loops becomes narrower but does not vanish for high $\tau$ values which is only possible for enhanced ferromagnetic core coupling values such as $J_{c}/J_{sh}=0.5$.

In the following analysis, let us investigate the effect of antiferromagnetic $J_{int}/J_{sh}$ interactions which can not be predicted from the results presented in figure \ref{fig7}. By comparing figures \ref{fig9}a and \ref{fig9}b we observe that for relatively strong antiferromagnetic interface couplings such as $J_{int}/J_{sh}=-2.5$ the system may exhibit an interesting phenomenon. Namely, as shown in figure \ref{fig9}b, the system exhibits triple hysteresis loops with an apparently wide middle loop for relatively strong $J_{int}/J_{sh}$ values. The width of the middle loop becomes wider as the field period $\tau$ increases. We note that observation of such hysteresis behavior is possible at sufficiently  low oscillation frequency values. More clearly, we have not observed this type of behavior in the system at oscillation periods $\tau\leq 1000$ with the parameters given in figure \ref{fig9}b. Origin of this phenomena can be understood by analyzing the time series of instantaneous magnetizations. According to our numerical calculations, triple hysteresis loop behavior originates from the existence of a weak ferromagnetic core coupling $J_{c}/J_{sh}$, as well as a strong antiferromagnetic interface exchange interaction $J_{int}/J_{sh}$. Triple loops disappear for $J_{int}/J_{sh}>0$. According to figure \ref{fig9}a where we consider a weak interface coupling $(J_{int}/J_{sh}=-0.5)$; at strong fields, such as for $h_{0}=-1.0$ both the core and shell layers can be magnetized along the field direction. As the field amplitude reaches to the value $h_{0}=1.0$ within a half cycle, both core and shell magnetizations easily keep their alignment with the external magnetic field. On the other hand, when $J_{int}/J_{sh}=-2.5$ (see figure \ref{fig9}b), shell magnetization can align in the oscillating field direction instantaneously, however core magnetization tends to align antiferromagnetically, due to the existence of a strong $J_{int}$ interaction and weak ferromagnetic core interaction $(J_{c}/J_{sh}=0.5)$. This results in a maximum phase lag between the magnetizations of the core and shell layers, and consequently we observe triple hysteresis loops. These types of hysteresis loops have also been observed recently in cylindrical Ising nanowire systems \cite{keskin}, and in molecular based magnetic materials \cite{jiang3} in the presence of static magnetic fields. However, triple hysteresis loops observed in the present system may be unphysical, since the interfacial interaction strength $J_{int}/J_{sh}$ is usually supposed to be within the range of $J_{c}<J_{int}<J_{sh}$ in the exchange bias systems \cite{iglesias3}.
\begin{figure}\begin{center}
\includegraphics[width=7.5cm]{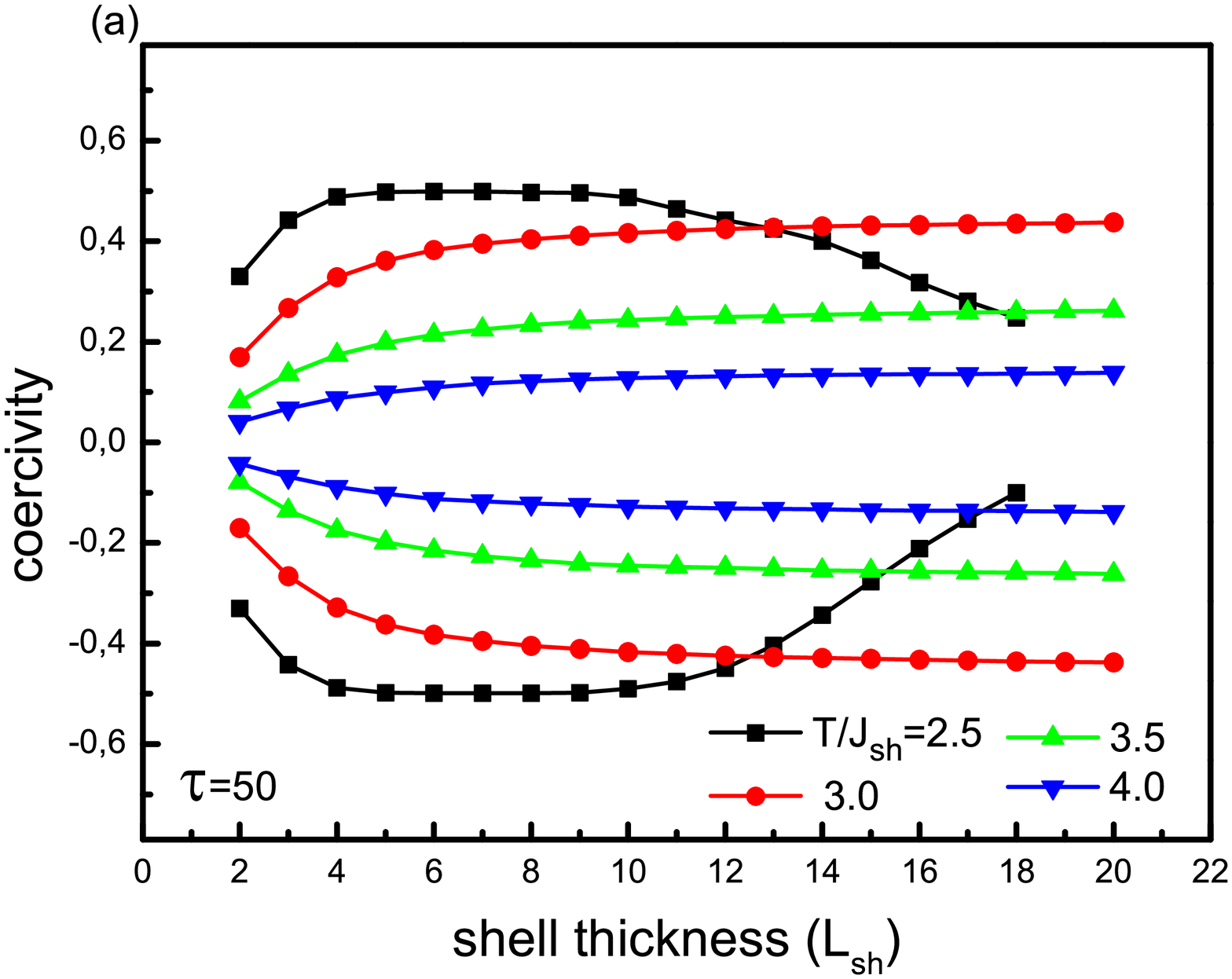}
\includegraphics[width=8.1cm]{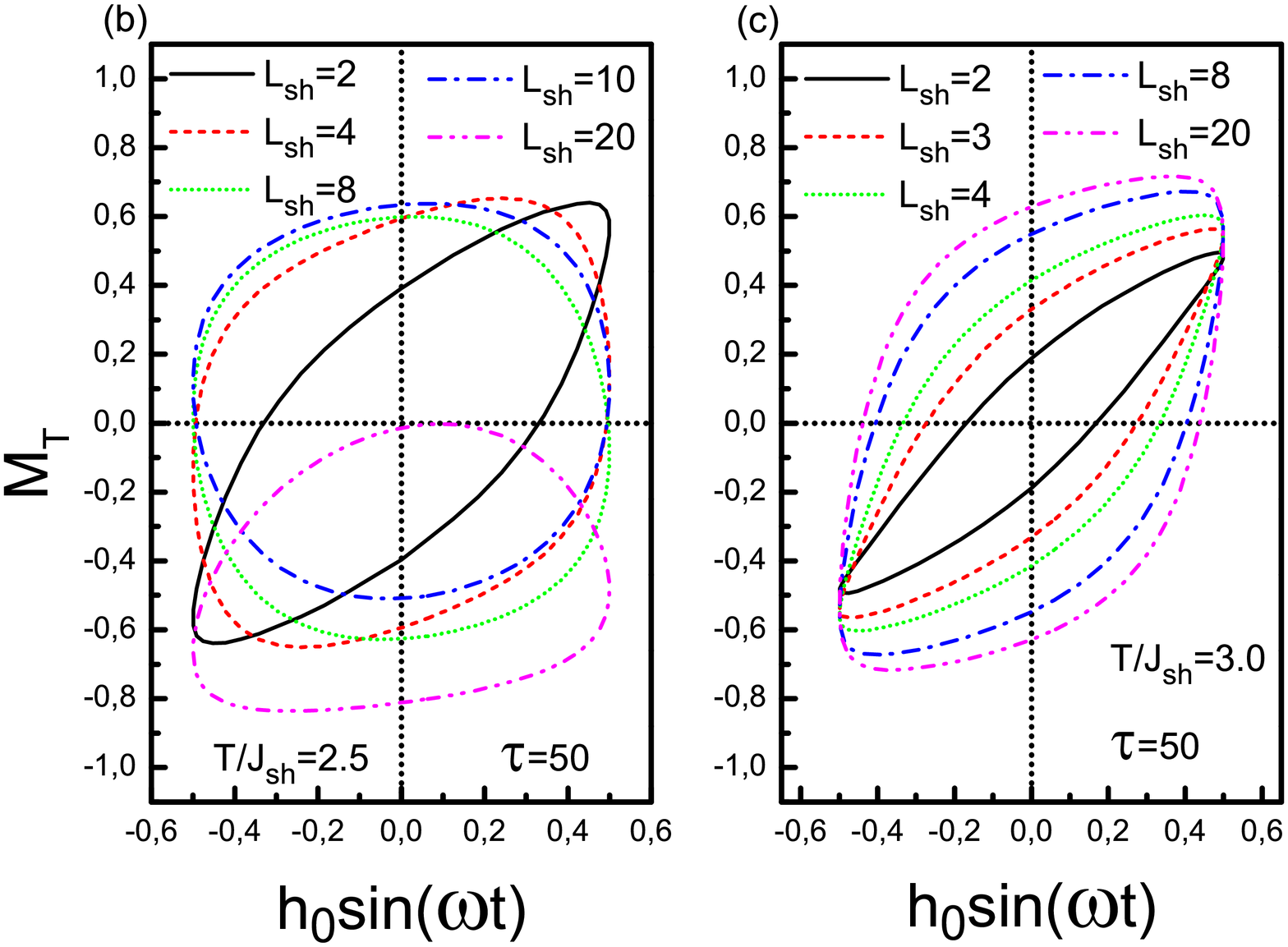}\\
\caption{(a) Dependence of the coercivity of the particle as a function of shell thickness $L_{sh}$ with some selected values of temperature. The system parameters have been kept fixed as $J_{int}/J_{sh}=-0.1$, $J_{c}/J_{sh}=0.25$, $\tau=50$, and $h_{0}=0.5$. Corresponding hysteresis curves of the total magnetization $M_{T}$ have been depicted for temperature values $T=2.5$ and $3.0$ in (b) and (c), respectively.}\label{fig10}
\end{center}
\end{figure}

\subsection{Size dependent properties}\label{results3}
\begin{figure}\begin{center}
\includegraphics[width=7.5cm]{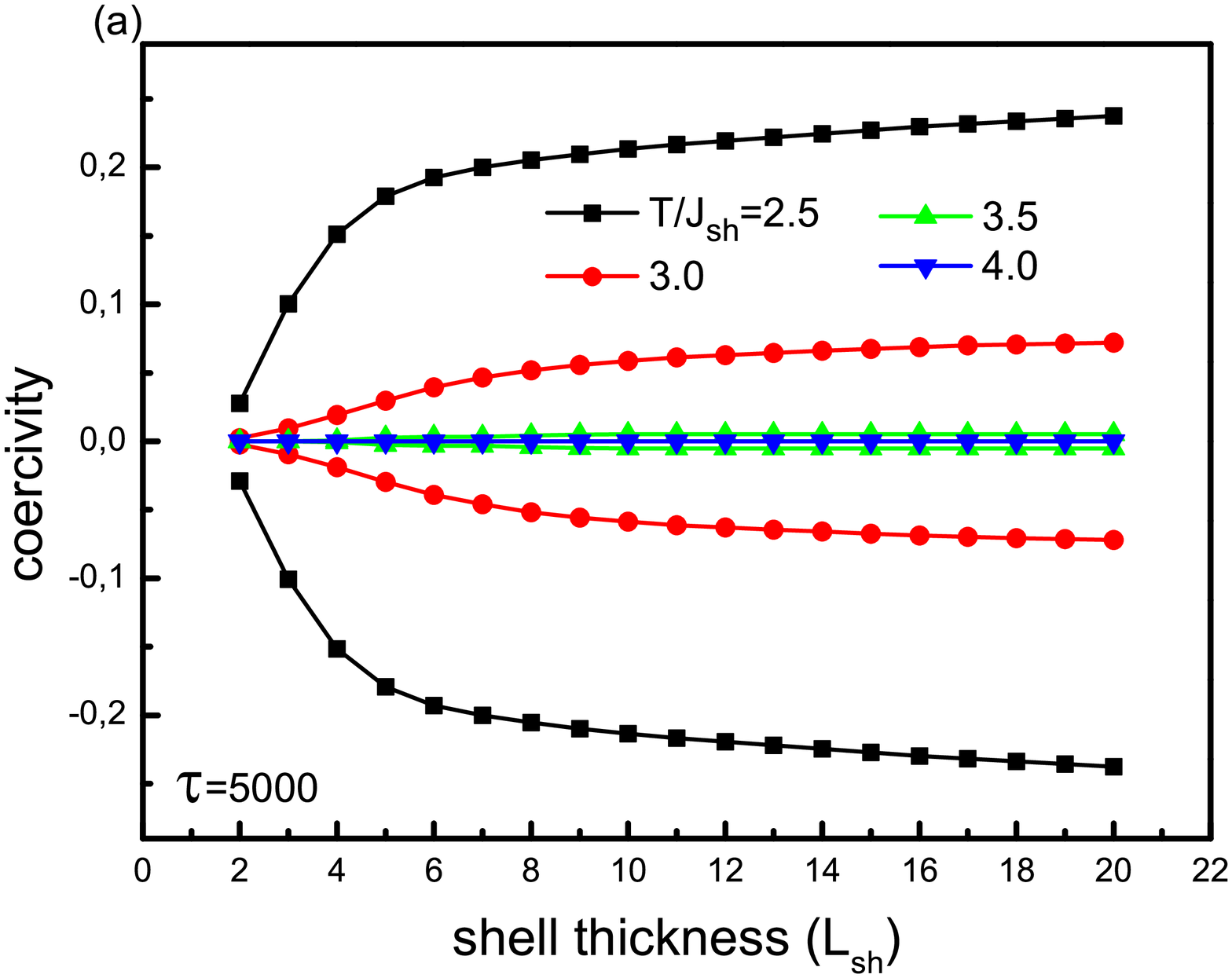}
\includegraphics[width=8.3cm]{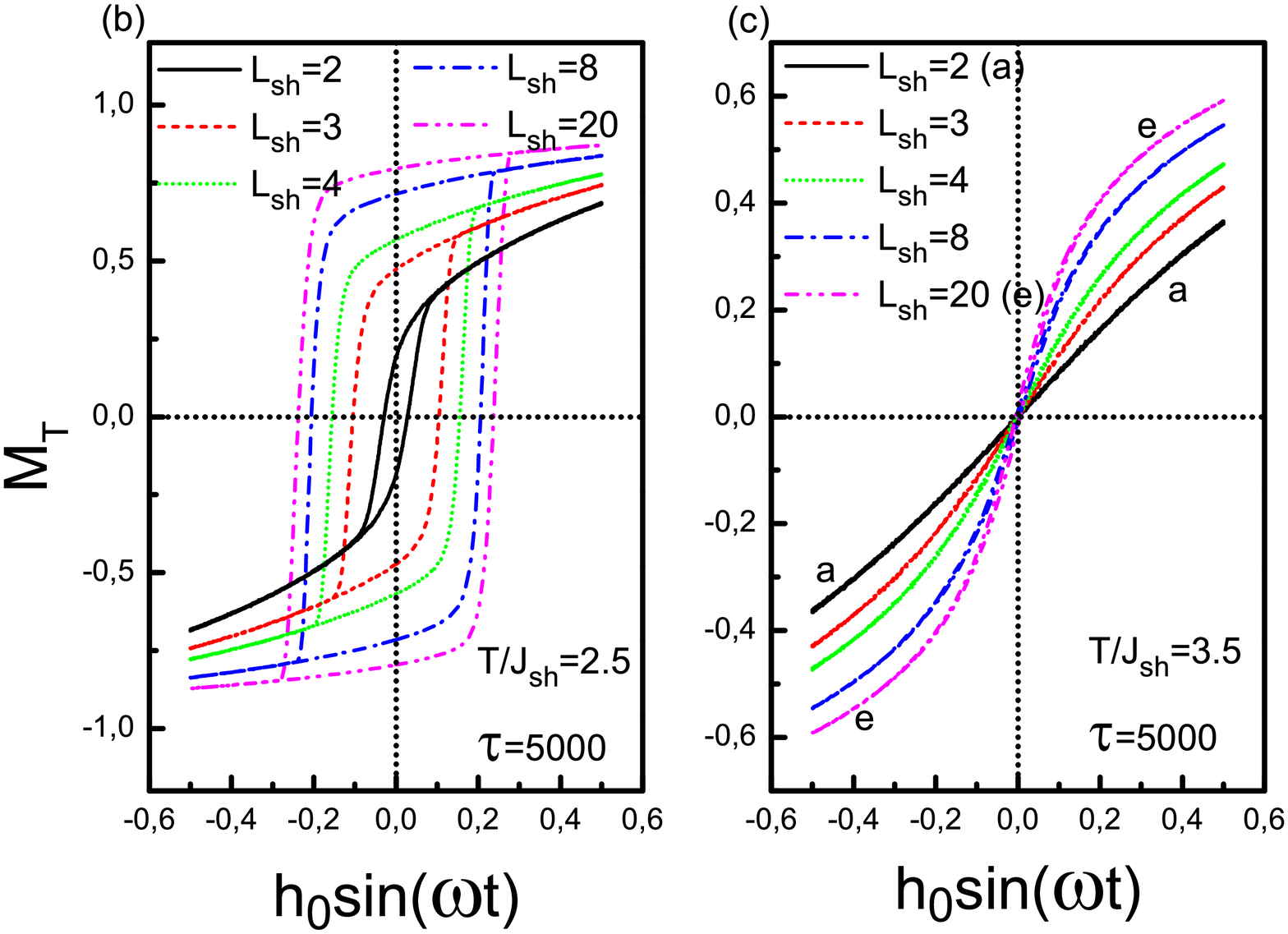}\\
\caption{(a) Dependence of the coercivity of the particle as a function of shell thickness $L_{sh}$ with some selected values of temperature. The system parameters have been kept fixed as $J_{int}/J_{sh}=-0.1$, $J_{c}/J_{sh}=0.25$, $\tau=5000$, and $h_{0}=0.5$. Corresponding hysteresis curves of the total magnetization $M_{T}$ have been depicted for temperature values $T=2.5$ and $3.5$ in (b) and (c), respectively.}\label{fig11}
\end{center}
\end{figure}

As a final investigation, let us represent some size dependent properties of the particle for some selected values of Hamiltonian and magnetic field parameters. In figure \ref{fig10}, we show the effect of the shell thickness $L_{sh}$ on the coercivity and hysteresis curves corresponding to high frequency regime $(\tau=50)$. It is clear from figure \ref{fig10}a that as the temperature increases then coercivity of the system increases and saturates at a certain value which depends on the temperature. Moreover, the curve corresponding to $T/J_{sh}=2.5$ (black full squares) exhibits an unusual behavior. Namely, coercivity of the system loses its symmetric shape with increasing $L_{sh}$ when $T/J_{sh}=2.5$ which can be regarded as a signal of a dynamic phase transition. Corresponding hysteresis curves have been depicted in figures \ref{fig10}b and \ref{fig10}c for $T/J_{sh}=2.5$ and $T/J_{sh}=3.0$, respectively. As seen from figure \ref{fig10}b, hysteresis loop loses its symmetry and the particle exhibits  a dynamic phase transition from paramagnetic to a dynamically ordered phase with increasing $L_{sh}$ values. This phase transition completely originates from the magnetization of the shell layer of the particle, since the magnetization of particle core does not change its shape as $L_{sh}$ varies. On the other hand, according to figure \ref{fig10}c, for $T/J_{sh}=3.0$ the particle always remains in the paramagnetic phase as $L_{sh}$ increases, since the temperature is large enough to keep the system in a dynamically disordered state. We can also observe from figure \ref{fig10}c that remanent magnetization values increase and loop areas get wider with increasing $L_{sh}$, and after a sufficiently large value of $L_{sh}$, the loop areas are not affected from changing $L_{sh}$ values. We have also investigated the situation for low frequency regime $(\tau=5000)$ in figure \ref{fig11}. At low frequency values, the magnetization of the particle is able to follow the external magnetic field instantaneously, but with a phase lag. Dependence of coercivity on the shell thickness $L_{sh}$ for $\tau=5000$ can be seen in figure \ref{fig11}a. By comparing figures \ref{fig10}a and \ref{fig11}a, we can clearly observe that the system exhibits large coercivities for high frequency values. Hysteresis curves corresponding to figure \ref{fig11}a are plotted in figures \ref{fig11}b and \ref{fig11}c for temperatures $T/J_{sh}=2.5$ and $T/J_{sh}=3.5$, respectively. The most remarkable observation in these figures is that the phase difference between the magnetization of the particle and oscillating external magnetic field drastically reduces, hence coercivity and remanent magnetization values become zero in low frequency and high temperature regions, and consequently loop areas reduce to zero.

\begin{figure}\begin{center}
\includegraphics[width=8.0cm]{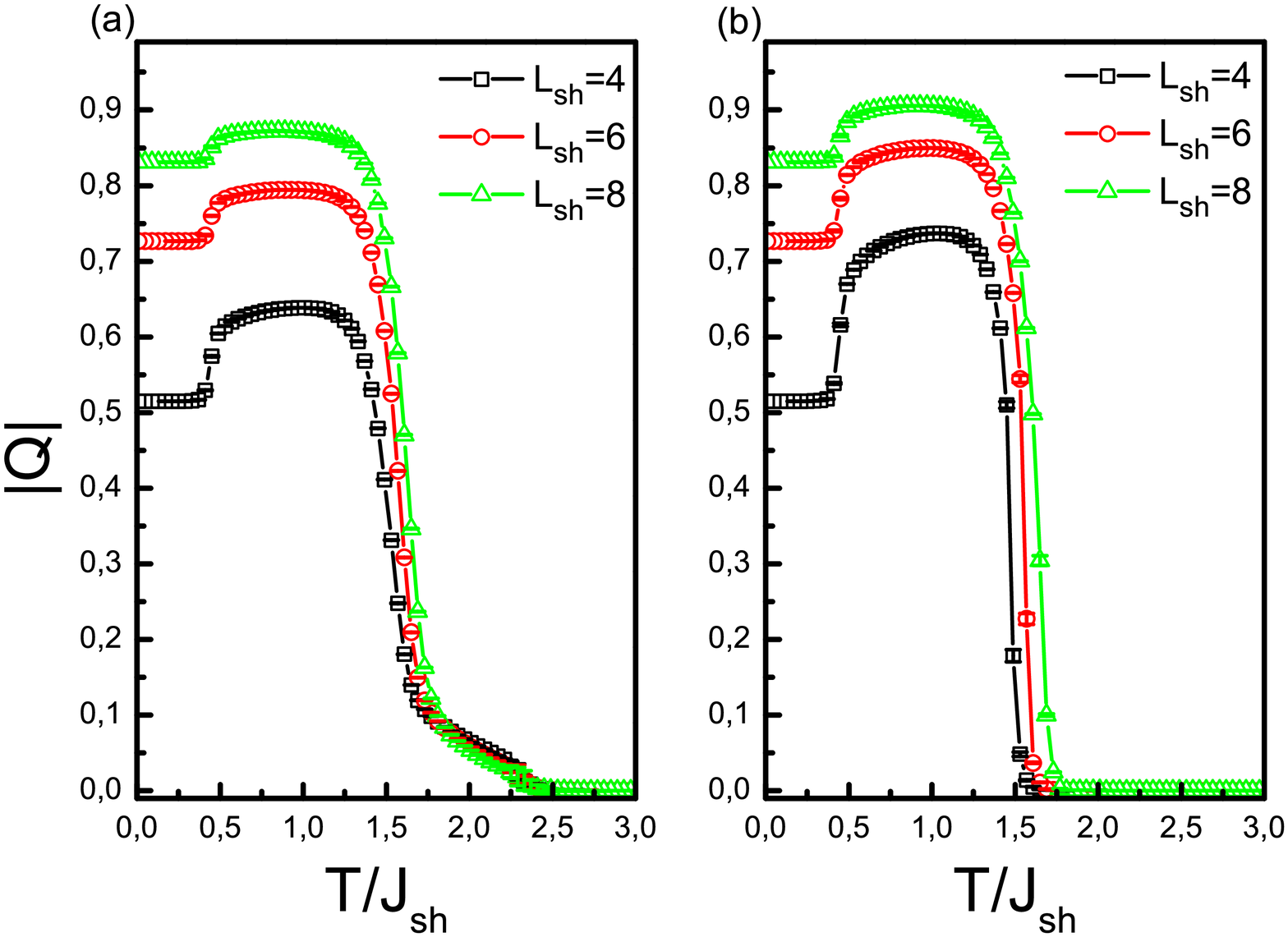}
\includegraphics[width=4.1cm]{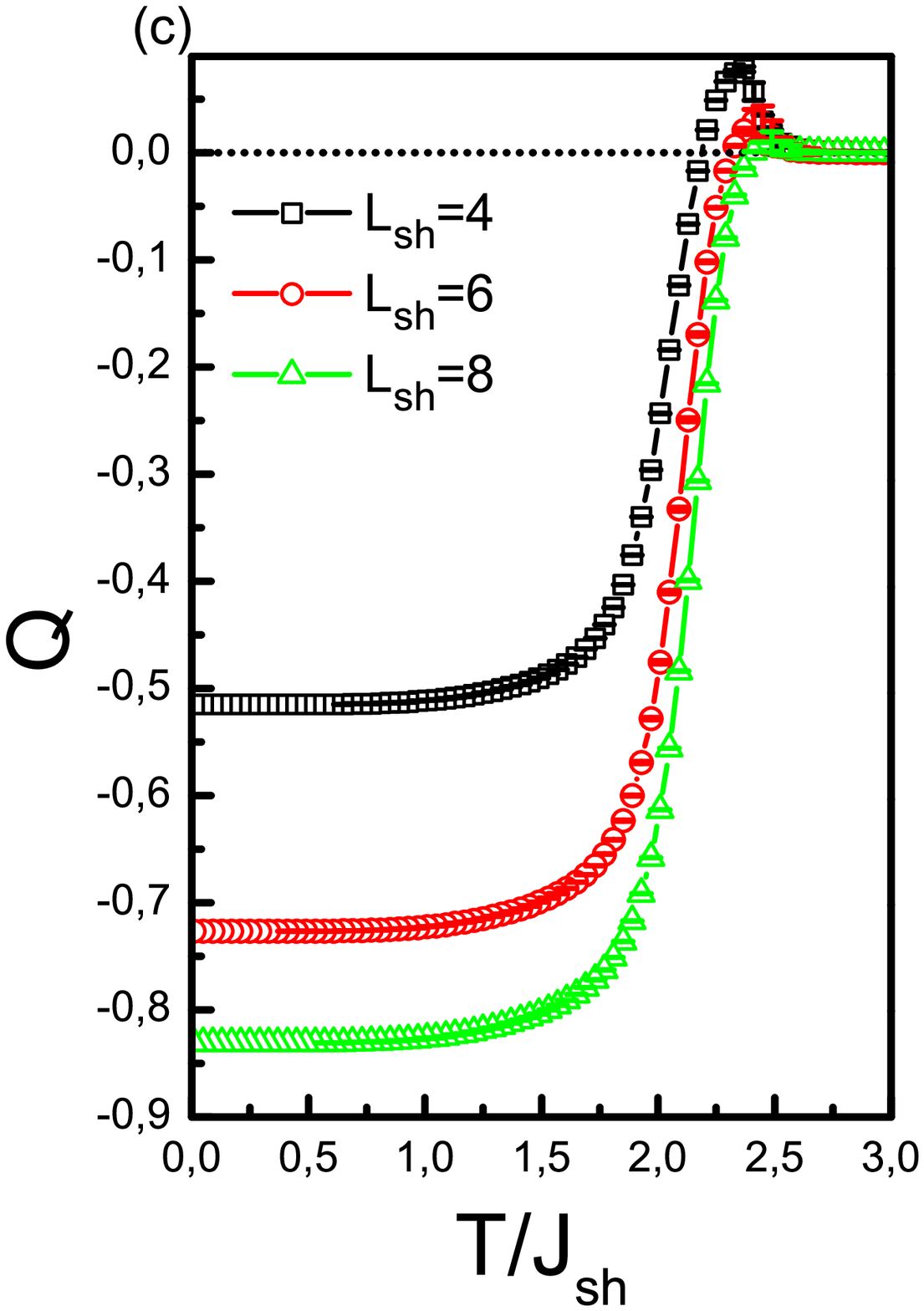}\\
\caption{Shell thickness $L_{sh}$ dependence of total magnetization $Q$ curves with a variety of system parameters. (a) $J_{int}/J_{sh}=-3.0$, $J_{c}/J_{sh}=0.25$, $h_{0}=1.0$, and $\tau=100$. (b) $J_{int}/J_{sh}=-0.5$, $J_{c}/J_{sh}=0.25$, $h_{0}=1.0$, and $\tau=50$. (c) $J_{int}/J_{sh}=-1.0$, $J_{c}/J_{sh}=0.6$, $h_{0}=0.75$, and $\tau=50$.}\label{fig12}
\end{center}
\end{figure}

In figures \ref{fig12}a-\ref{fig12}c, in order to depict the effect of the ferromagnetic shell layer thickness $L_{sh}$ on $T_{comp}$ and $T_{c}$ values, we plot the temperature dependencies of total magnetization curves of the particle for some selected values of $L_{sh}$. Figures \ref{fig12}a and \ref{fig12}b show the representative P-type curves with some selected values of Hamiltonian parameters and both for fairly strong, and relatively moderate antiferromagnetic interface couplings such as $J_{int}/J_{sh}=-3.0$ and $-1.0$, respectively. It is clear from figures \ref{fig12}a and \ref{fig12}b that although the system exhibit a ferrimagnetic order, we cannot observe any compensation point. Moreover, it can be easily observed in figures \ref{fig12}a and \ref{fig12}b that the temperature induced maxima of the total magnetization curves get higher with increasing $L_{sh}$, however the magnetization curves retain their shapes even if the particle size changes and the curves approach to bulk limit with increasing $L_{sh}$ values. On the other hand, when the surface exchange enhancement is somewhat reduced, such as $J_{c}/J_{sh}=0.6$ case depicted in Fig. \ref{fig12}c, the total magnetization curves exhibit N-type behavior with a compensation point for thinner shell layers. As the shell thickness increases, then the total magnetization curves exhibit a shape transformation from N-type characteristics to ordinary Q-type shape where we observe a monotonic decrease in the magnetization with  increasing temperature.

\section{Concluding remarks}\label{conclude}
In conclusion, by means of MC simulations, in order to clarify how the magnetism in a nanoparticle system is affected in the presence of a periodically  oscillating external perturbation,  we have analyzed nonequilibrium phase transition properties and stationary-state behavior of a single domain ferrimagnetic nanoparticle which is composed of a ferromagnetic core surrounded by a ferromagnetic shell layer. By considering an antiferromagnetic exchange interaction in the interfacial region we have also investigated some of the ferrimagnetic properties of the particle. The most conspicuous observations reported in the present paper can be briefly summarized as follows:

\begin{itemize}
\item
In Section \ref{results1}, a complete picture of the phase diagrams and magnetization profiles have been presented. We have observed that in the presence of surface exchange enhancement (such as $J_{c}/J_{sh}=0.25$) and at high oscillation frequencies the system does not exhibit compensation phenomena. Moreover, we have found that the existence of strong antiferromagnetic exchange interaction (such as $J_{int}/J_{sh}=-2.5$) is not sufficient for the occurrence of compensation phenomena in the system. $J_{c}/J_{sh}$ parameter is the decisive factor for the occurrence of a compensation point. In other words, there exists a critical $J_{c}/J_{sh}$ value below which the system can not exhibit compensation point. Compensation points have been found to reduce as external field amplitude increases. According to N\'{e}el nomenclature, the magnetization curves of the particle are found to obey  P-type, N-type and Q-type classification schemes under certain conditions. The numerical values of dynamic magnetizations, especially the total magnetization $Q$ of the system is defined in such a way that different number of spins in the core and shell layers of the particle are considered. In this context, recent calculations reported in the literature should be treated carefully, since the number of spins in the core and shell are generally not equal in the core-shell nanoparticle models.
\item
Section \ref{results2} has been devoted to investigation of hysteretic response of the particle. Based on the simulation results, we have reported the existence of triple hysteresis loop behavior which originates from the existence of a weak ferromagnetic core coupling $J_{c}/J_{sh}$, as well as a strong antiferromagnetic interface exchange interaction $J_{int}/J_{sh}$. Triple hysteresis loops disappear under the transformation $J_{int}\rightarrow -J_{int}$. However, we claim that triple hysteresis loops observed in the present system may be unphysical, since the interfacial interaction strength $J_{int}/J_{sh}$ is usually supposed to be within the range of $J_{c}<J_{int}<J_{sh}$ in the exchange bias systems \cite{iglesias3}
\item Size dependent properties of the particle have been clarified in Section \ref{results3} and we found that the particle may exhibit a dynamic phase transition from paramagnetic to a dynamically ordered phase with increasing ferromagnetic shell thickness $L_{sh}$ when the oscillation frequency is sufficiently high which simulates an highly nonequilibrium scenario in the presence of ultrafast fields. We have also found that as the shell thickness increases, then the total magnetization curves exhibit a shape transformation from N-type characteristics to ordinary Q-type, hence compensation point may disappear with increasing $L_{sh}$.
\end{itemize}

All of the observations outlined above show that the shape (amplitude and frequency) of the driving field and particle size have an important influence on the thermal and magnetic properties, such as coercivity, remanence and compensation temperature of the particle. In fact, we note that according to our simulation results, we have not found any evidence of the first order phase transitions. The reason is most likely due to the fact that, in contrast to the conventional techniques such as MFT and EFT, the method we used in the present work completely takes into account the thermal fluctuations in the present system which allows us to obtain non-artificial results. Moreover, it is possible to improve the proposed model to simulate more realistic systems by considering a simulation of Heisenberg type of Hamiltonian with an assembly of interacting nanoparticles instead of a single particle. This may be the subject of a future work.

On the other hand, we should also note that the conventional MC modeling of dynamical systems is capable of explaining the time dependent properties, such as dynamic order parameters in terms of an artificial timescale. Recently, in order to relate the dynamic parameters to any real time scale, a nice theoretical attempt has been introduced as a dynamical approach using the Landau-Lifshitz-Gilbert (LLG) equation \cite{nowak,cheng2, cheng3, hinzke, hinzke2, rueda, hinzke3, berkov}. The method which was propounded for the first time by Nowak \emph{et al.} \cite{nowak} is simply based on the introduction of a time quantification factor which relates MC time step to physical time used in the LLG equation. This method is widely regarded to be suitable for modeling short-time  scale dynamics where the time step is only of the order of several picoseconds \cite{cheng}. Hence, it could also be interesting to treat the problem presented in this study within the framework of a time-quantified Monte Carlo technique.

\section*{Acknowledgements}
The authors (Y.Y. and E.V.) would like to thank the Scientific and Technological Research Council of Turkey (T\"{U}B\.{I}TAK) for partial financial support.
This work has been completed at Dokuz Eyl\"{u}l University, Graduate School of Natural and Applied Sciences, and the numerical calculations reported in
this paper were performed at T\"{U}B\.{I}TAK ULAKBIM, High Performance and Grid Computing Center (TR-Grid e-Infrastructure).


\end{document}